\documentstyle[preprint,aps]{revtex}
\draft
\begin{document}
\newcommand{\eq}[1]{eq.~(\ref{#1})}
%\twocolumn[
\bibliographystyle{prsty}
\preprint{RUB-TPII-07/96, hep-ph/9609222 }
\title{Weak electricity of  the Nucleon \\ in the
Chiral Quark-Soliton Model}
\author{
Hyun-Chul Kim
\footnote{Electronic address: kim@hadron.tp2.ruhr-uni-bochum.de},
Maxim V. Polyakov
\footnote{Electronic address: maximp@hadron.tp2.ruhr-uni-bochum.de}
\footnote{Alexander von Humboldt Fellow,
on leave of absence from Petersburg Nuclear Physics
Institute, Gatchina, St. Petersburg 188350, Russia},
and Klaus Goeke
\footnote{Electronic address: goeke@hadron.tp2.ruhr-uni-bochum.de}
}
\address{
Institute for  Theoretical  Physics  II, \\
Ruhr-University Bochum, \\
 D-44780 Bochum, Germany        }

\date{September, 1996    }
\maketitle
%\widetext
\begin{abstract}
The induced pseudotensor constant (weak electricity)
of the nucleon is  calculated in the framework of the chiral quark
soliton model.  This quantity originates from the G--parity violation
and hence is proportional to $m_u-m_d$.
We obtain  for $m_u-m_d=-5$~MeV a value of  $g_T/g_A =-0.0038$.
\end{abstract}

%]
%\narrowtext
%\vfill\eject
\section{Introduction}

The neutron $\beta$-decay is a powerful tool to probe
the structure of the nucleon.  In particular, it provides
a precise measurement of the triplet axial constant of the nucleon $g_A$,
so that it is a touchstone for any model of the nucleon structure.
The underestimation of the nucleon axial charge in the solitonic
 picture of the nucleon was for a long  main critical point of
soliton models of the nucleon.  Recently it was shown
\cite{WakWat,Chretal} that in the chiral quark-soliton model of the
nucleon ($\chi$QSM) the rotational $1/N_c$ corrections to the $g_A$
bring its value close to the experimental one.  Also these corrections
improve considerably the agreement of the electromagnetic
characteristics of baryons \cite{ChrGorGoePo,KimPolBloGo,KimBloPolGo}
calculated in the $\chi$QSM with an experiment.

In the present paper we investigate the other than $g_A$
 axial characteristic of the
nucleon -- induced pseudotensor constant (weak electricity) of the
nucleon $g_T$.  The neutron-to-proton transition matrix element of the
axial current $J^5_\mu=\bar u \gamma_\mu \gamma_5 d$ can be written in
terms of three form factors:
\begin{equation}
\langle P(p^\prime) | J^5_{\mu} |N(p)\rangle= \bar{u}_p(p^\prime)
\bigl \{g_A\gamma_{\mu} \gamma_5
+ \frac{g_T}{M_p+M_n}i \sigma_{\mu \nu}\gamma_5 q_{\nu}
+g_P q_{\mu} \gamma_5 \bigr\} u_n(p)\,, \qquad q=p^\prime-p,
\label{matrele}
\end{equation}
where $M_p$ ($M_n$) is the proton (neutron) mass and we use the
convention of Bjorken and Drell for Dirac matrices and spinors.  The
axial-vector $g_A$ and pseudoscalar $g_P$\footnote{The pseudoscalar
axial constant $g_P$ was calculated recently in ref.~\cite{WatChrGoe}
in the framework of the chiral quark soliton model}
 constants were extensively analyzed theoretically and measured
in experiments, while less is known about the pseudotensor constant
$g_T$.  The pseudotensor current has the opposite $G$-parity  to that
of the axial vector current and hence is proportional to the parameter
of isospin symmetry breaking.  There are two different sources of
isospin symmetry breaking:  Electromagnetic interactions and $u$ and
$d$ quark mass difference.  In this work we calculate the hadronic part
of the $g_T$ proportional to $m_u-m_d$ in the limit of a large number
of colors, $N_c \rightarrow\infty$.

Even though in reality $N_c=3$,
the limit of large $N_c$ furnishes a useful guideline.
At large $N_c$ the nucleon is heavy and can
be viewed as a classical soliton of the pion field \cite{Witten}. An
example of the dynamical realization of this idea is  given by the
Skyrme model~\cite{ANW}.
A far more realistic effective chiral lagrangian of the $\chi$QSM
is based on the interaction of dynamically
massive constituent quarks with pseudo-Goldstone meson fields.  It is
given by the functional integral over the quark ($\psi$) in the background
pion field \cite{DE,ATFR,DSW,DyPe1}:
 \begin{equation}
\exp\left(iS_{\rm eff}[\pi(x)]\right)\; = \; \int {\cal D}\psi
{\cal D}\bar\psi
\exp{\left(i \int d^4 x \bar\psi D \psi \right)},
\label{eq:Z}
\end{equation}
where $D$ is the Dirac operator
\begin{equation}
 D \; = \;  i \rlap{/}{\partial} -\hat{m} -MU^{\gamma_5}.
\label{Dirac_operator}
\end{equation}
$U^{\gamma_5}$ denotes the pseudoscalar chiral field
\begin{equation}
U^{\gamma_5}\;=\;\exp{i\pi^a \tau^a \gamma_{5}}
\;=\;\frac{1+\gamma_5}{2}U\;+\;\frac{1-\gamma_5}{2} U^{\dagger}.
\end{equation}
The $\hat{m}$ is the matrix of the current quark masses
$\hat{m}\;=\;\mbox{diag}(m_u, m_d)$.
The $M$ stands for the dynamical quark mass arising as a
result of the spontaneous chiral symmetry breaking.

The effective chiral action given by \eq{eq:Z} is known to
contain automatically the Wess--Zumino term and the four-derivative
Gasser--Leutwyler terms, with correct coefficients. Therefore, at least the
first four terms of the gradient expansion of the effective chiral
lagrangian are correctly reproduced by \eq{eq:Z}, and chiral symmetry
arguments do not leave much room for further modifications.
Eq.~(\ref{eq:Z}) has been derived from the instanton model of the QCD
vacuum \cite{DyPe1}, which provides a natural mechanism of chiral
symmetry breaking and enables one to express the dynamical mass $M$ and the
ultraviolet cutoff $\Lambda$ intrinsic in \eq{eq:Z} through the
$\Lambda_{QCD}$ parameter. It should be mentioned that \eq{eq:Z} is
of general nature: one can use \eq{eq:Z} without referring to
the instantons.

An immediate implication of the effective chiral theory \eq{eq:Z} is the
quark-soliton model of baryons \cite{DiPePo}.
According to these ideas the nucleon
can be viewed as a bound state of $N_c$ (=3) {\em valence}
quarks kept together
by a hedgehog-like pion field whose energy coincides by definition
with the aggregate energy of quarks from the negative Dirac sea.  Such a
semiclassical picture of the nucleon is well justified in the limit
$N_c\rightarrow\infty$ -- in line with more general arguments by Witten
\cite{Witten}.
The further studies showed that the $\chi$QSM is
successful in reproducing the static properties
and form factors of the baryons using just one parameter set
(see the recent review \cite{Review}). The powerful numerical method
to carry out the calculation of the $N-\Delta$ splitting and nucleon
matrix elements of arbitrary quark bilinear operators has been
developed in refs.~\cite{WakYos,Goeetal}. This method is also used in
the present paper.

\section{Computing weak electricity}

The transition matrix element eq.~(\ref{matrele})
can be computed as the Euclidean functional integral in the $\chi$QSM
\begin{eqnarray}
\langle P |\bar u \gamma_\mu\gamma_5 d | N \rangle & = &
\frac{1}{\cal Z} \lim_{T \rightarrow \infty} \exp{(ip_0 \frac{T}{2}
- ip'_{0} \frac{T}{2})} \nonumber \\
& \times & \int d^3 x d^3 y
\exp{(-i \vec{p'} \cdot \vec{y} + i \vec{p} \cdot \vec{x})}
\int {\cal D}U \int {\cal D} \psi \int {\cal D}\psi^\dagger
\nonumber \\
& \times & \; J_{p}(\vec{y},T/2)\, \bar u \gamma_\mu
\gamma_5 d\, J^{\dagger}_{n} (\vec{x}, -T/2)
%\nonumber \\ & \times &
\exp{\left[\int d^4 z \psi^\dagger D \psi \right ]}.
\label{Eq:ev}
\end{eqnarray}
The nucleon current $J_N$ ($N=p,n$) is
built of $N_c$ quark fields:
\begin{equation} J_N(x)\;=\; \frac{1}{N_c !}
\epsilon_{i_1 \cdots i_{N_c}} \Gamma^{\alpha_1 \cdots
\alpha_{N_c}}_{JJ_3TT_3}\psi_{\alpha_1i_1}(x)
\cdots \psi_{\alpha_{N_c}i_{N_c}}(x).
\end{equation}
$\alpha_1 \cdots\alpha_{N_c}$ denote spin--flavor indices, while
$i_1 \cdots i_{N_c}$ designate color indices.  The matrices
$\Gamma^{\alpha_1 \cdots\alpha_{N_c}}_{JJ_3TT_3}$ are taken to endow
the corresponding current with the quantum numbers $JJ_3TT_3$.

In the large $N_c$ limit the integral over Goldstone fields $U$ in
eq.~(\ref{Eq:ev}) can be calculated by the steepest descent method
(semiclassical approximation).  The corresponding saddle point equation
admits a static soliton solution, an example of which is the hedgehog
field configuration:
\begin{equation}
U_s(\vec x)\;=\;\exp{[i\vec{n}\cdot\vec{\tau}P(r)]} .
 \label{ezh}
\end{equation}
The $P(r)$ denotes the profile function satisfying the boundary condition
$P(0)=\pi$ and $P(\infty)=0$, which is determined by solving the
saddle point equations (for details see Ref.~\cite{Review}).
The soliton is quantized by introducing collective coordinates
corresponding to $SU(2)_{I}$ isospin rotations of the soliton
(and simultaneously $SU(2)_{spin}$ in spin space):
\begin{equation}
U(t,\vec{x})=R(t)U_s(\vec{x})R^\dagger(t),
\end{equation}
where $R(t)$ is a time--dependent $SU(2)$ matrix.
The quantum states arising
from this quantization have the quantum numbers corresponding to
the nucleon and $\Delta$.

Calculating the functional integral eq.~(\ref{Eq:ev}) we obtain the
following expression for the neutron to proton transition element of
the  axial current:
\begin{eqnarray}
\nonumber
\langle P |\bar u \gamma_\mu\gamma_5 d | N \rangle&=&
N_c(M_p+M_n) \int d^3x e^{i{\bf q\cdot x}}\int dR \, \phi_p^\ast(R)  \\
&\times& \int \frac{d\omega}{2\pi} \mbox{\rm tr}\Biggl(
\langle {\bf x} | \,
\frac{1}{\omega + i H+
 i(m_u-m_d)R^\dagger \tau^3 R}\,\gamma_0 \gamma_\mu \gamma_5 \,
R^\dagger \tau^{1+i2}R
|{\bf x}\rangle
\Biggr)\, \phi_n(R),
\label{axial:trace}
\end{eqnarray}
where $\phi^{S=T}_{S_3T_3}(R)$ is the rotational wave function of the
nucleon ($\phi_p=\phi^{(\frac 12)}_{\frac 12 \frac 12}$,
 $\phi_n=\phi^{(\frac 12)}_{\frac 12 -\frac 12}$) given by the
 Wigner finite-rotation matrix \cite{ANW,DiPePo}:

\begin{equation}
\phi^{S=T}_{S_3T_3}(R) =
\sqrt{2S+1}(-1)^{T+T_3}{\cal D}^{S=T}_{-T_3,S_3}(R),
\label{Wigner}
\end{equation}
and the integral over $SU(2)$ group is normalized by $\int dR =1$.
The one-particle Dirac Hamiltonian $H$ in a background of the static pion
field eq.~(\ref{ezh}) has a form
 \begin{equation}
H= \gamma^0\gamma^k \partial_k +i  M\gamma^0 U_s^{\gamma_5}
+\frac12 (m_u+m_d)\, .
\label{Ham}
 \end{equation}

Projecting the general expression eq.~(\ref{axial:trace}) onto the
pseudotensor structure one obtains:

\begin{eqnarray}
\nonumber
\frac{g_T(q^2)}{M_p+M_n}&=&
N_c \int d^3x e^{i{\bf q x}}\frac{q^3}{|{\bf q}|^2}
\int dR \, \phi_p^\ast(R)  \\
&\times& \int \frac{d\omega}{2\pi} \mbox{\rm tr}\Biggl(
\langle {\bf x} | \,
\frac{1}{\omega + i H+
 i(m_u-m_d)R^\dagger \tau^3 R}\,\gamma_5 \,
R^\dagger \tau^{1+i2}R
|{\bf x}\rangle
\Biggr)\, \phi_n(R).
\label{g2:trace}
\end{eqnarray}

Let us now show that the above expression is zero  in the isospin
symmetry limit ($m_u=m_d$). To prove this we introduce the following
unitary transformation of the Dirac and Pauli matrices connecting
them to the transposed ones:
\begin{equation}
W\gamma_\mu W^{-1}=\gamma_\mu^T, \quad
W \tau^a W^{-1}=-(\tau^a)^T.
\end{equation}
Evidently then $W H W^{-1}=H^T$.
Using properties of the trace $\mbox{tr}(M^T)=\mbox{tr}(M)$ and
$\mbox{tr}(WMW^{-1})=\mbox{tr}(M)$
one can write:
\begin{eqnarray}
\nonumber
&&\mbox{\rm tr}\Biggl(\langle {\bf x} | \,
\frac{1}{\omega + i H+
 i(m_u-m_d)R^\dagger \tau^3 R}\,\gamma_5 \,
R^\dagger \tau^{1+i2}R
|{\bf x}\rangle
\Biggr)= \\
\nonumber
&&\mbox{\rm tr} \Biggl(W\langle {\bf x} | \,
\frac{1}{\omega + i H+
 i(m_u-m_d)R^\dagger \tau^3 R}\,\gamma_5 \,
R^\dagger \tau^{1+i2}R
|{\bf x}\rangle W^{-1}
\Biggr)^T= \\
\nonumber
&-&\mbox{\rm tr}\Biggl(\langle {\bf x} | \,
\frac{1}{\omega + i H-
 i(m_u-m_d)R^\dagger \tau^3 R}\,\gamma_5 \,
R^\dagger \tau^{1+i2}R
|{\bf x}\rangle
\Biggr)\,.
\label{g2:zero}
\end{eqnarray}
This immediately implies that the pseudotensor constant given by
eq.~(\ref{g2:trace}) is zero in the isosymmetric limit and first
non-zero result appears expanding eq.~(\ref{g2:trace})
in $m_u-m_d$ to linear order.
The result for the pseudotensor constant $g_T$ in the leading order of
$1/N_c$  expansion ($g_T\sim N_c$) and the linear order in $m_u-m_d$
has a form \cite{ich}:

 \begin{eqnarray}
 \frac{g_T}{M_p+M_n}&=&
\frac{i N_c(m_u-m_d)}{24}
 \int \frac{d \omega}{2 \pi} \mbox{\rm Sp}(
\frac{1}{\omega + i H }\,\gamma_0 \tau^i\, \frac{1}{\omega + i H }
\,\varepsilon_{ijk }\tau^j x^k \gamma_5 ) \nonumber \\
&\times& \varepsilon_{ab3}\, \int dR\,
\phi_p^\ast(R)
 {\cal D}_{1+i2,a}^{(1)}(R)
 {\cal D}_{3,b}^{(1)}(R)
\phi_n(R) \, .
 \label{g2}
 \end{eqnarray}
The  integral over soliton orientations in the second line of
 eq.~(\ref{g2}) can be easily calculated by using the relations
 \begin{equation}
 \varepsilon_{ab3}\,
 {\cal D}_{1\pm i2,a}^{(1)}(R) {\cal D}_{3,b}^{(1)}(R)  =
 \pm i {\cal D}_{1\pm i2,3}^{(1)}(R),
\end{equation}
 and
 \begin{equation}
\int dR\,
\phi_p^\ast(R)
 {\cal D}_{1+i2,3}^{(1)}(R)
\phi_n(R)                   = -2/3.
\label{kin:part}
\end{equation}
The functional trace in the first line of eq.~(\ref{g2}) was estimated
in ref.~\cite{ich} by means of the gradient expansion:
\begin{eqnarray}
\nonumber
\frac{g_T}{M_p+M_n}&\approx&
\frac{ N_c(m_u-m_d)}{9\cdot96\pi^2 M}\mbox{\rm Im } \int d^3 x \,
\Bigl[\varepsilon_{klm}
\mbox{\rm tr }\bigl( U \partial_k U^\dagger\partial_l U \tau^m  \bigr)
\\&-&
\nonumber
\frac i4 \varepsilon_{klm}\varepsilon^{abm}
\mbox{\rm tr }\bigl(\tau^b
\partial_kU(\tau^aU-U^\dagger\tau^a)\partial_lU^\dagger
\bigr)
\\
\nonumber
&-&
\frac i2 x^i\varepsilon_{klm}\varepsilon^{abi}
\mbox{\rm tr }\bigl(\tau^b U^\dagger
\partial_kU(\tau^a\partial_l U-\partial_l U^\dagger\tau^a)U^\dagger
\partial_m U
\bigr)
\\
&-&
\frac i2 x^i\varepsilon_{klm}\varepsilon^{abi}
\mbox{\rm tr }\bigl((\tau^b U^\dagger \tau^a-\tau^a U^\dagger \tau^b)
\partial_kU^\dagger \partial_l U\partial_m U^\dagger U\bigr)
\Bigr] \, .
\label{grad:exp}
\end{eqnarray}
This approximation is justified only for a soliton of large size
$RM\gg1$.
The real nucleon has a radius of order $1/M$ and hence
the eq.~(\ref{grad:exp}) can be used only as an order of
magnitude estimate.

\section{Numerical results and conclusion}

In order to evaluate exactly
the functional trace in eq.~(\ref{g2}),
we diagonalize the Hamiltonian $H$ eq.~(\ref{Ham}) numerically
in the Kahana-Ripka discretized basis~\cite{KaRi}.
The constituent quark mass $M$ is fixed to 420 MeV
in our model by reproducing best many static baryon observables
and form factors in the model (in particular, the isospin mass
splittings for octet and decuplet baryons  \cite{PraBloGo1,Review}).
To make sure of the
 numerical calculation, we compare our results for $g_T$ with the
analytical ones of the gradient expansion eq.~(\ref{grad:exp})
justified in the limit of large soliton size.  Our numerical procedure
is in good agreement within a few percent with the analytical
results of the gradient expansion in the large soliton size limit.

The results of our calculation are summarized in Table~I. For
completeness  we give in Table~I  also results for $g_A$ obtained
in \cite{Chretal}.  Let us note that the present result is comparable
to a recent calculation of the nucleon pseudotensor constant with the QCD
sum rule technique \cite{Shiomi} which gives $g_T/g_A=-0.0151 \pm 0.0053$.
 Both the QCD sum rule result and ours are in agreement with the bag
model calculation ($g_T/g_A=-0.00455$)\cite{Shiomi,DoHo}, whereas they
seem to be smaller than preliminary experimental data \cite{Morita}
ranging from $-0.21 \pm 0.14 $ to $0.14 \pm 0.10$.  However, the
accuracy of the experiment is not enough to be compared in a reasonable
way  with the results of theoretical calculations.

\section*{Acknowledgments}
This work has partly been supported by the BMBF, the DFG
and the COSY--Project (J\" ulich).  The work of M.V.P. is supported
by the Alexander von Humboldt Foundation.
\begin{table}
\caption{ Axial vector $g_A$ and pseudotensor $g_T$ constants of
the nucleon as a function of the constituent quark mass $M$,
$m_u-m_d=-5$~MeV.}
\begin{tabular}{ccc}
$M$~[MeV]& $g_A$ \protect\cite{Chretal}   &$ g_T/g_A$  \\
\hline
     370 & 1.26  &  -0.0029  \\
     400 & 1.24  &  -0.0035  \\
     420 & 1.21  &  -0.0038  \\
     450 & 1.16  &  -0.0040  \\
\end{tabular}
\end{table}

\vfill\eject
\vfill\eject
%====================================================================
\newpage
\pagebreak

\begin{thebibliography}{99}
\bibitem{WakWat} M. Wakamatsu and T. Watabe, {\em Phys. Lett.}
{\bf 312B}, 184 (1993).
\bibitem{Chretal} C.V. Christov, A. Blotz, K. Goeke,
P. Pobylitsa, V. Petrov, M. Wakamatsu, and T. Watabe,
{\em Phys. Lett.} {\bf 325B}, 467 (1994).
\bibitem{ChrGorGoePo} C.V. Christov, A.Z. G\'orski, K. Goeke, and
P. Pobylitsa, {\em Nucl. Phys.}, {\bf A592}, 513 (1995).
\bibitem{KimPolBloGo}
 H.-C. Kim, M.V. Polyakov, A. Blotz, and
K. Goeke, {\em Nucl. Phys.} {\bf A598}, 379 (1996).
\bibitem{KimBloPolGo}
 H.-C. Kim,  A. Blotz, M.V. Polyakov, and
K. Goeke, {\em Phys. Rev.} {\bf D53}, 4013 (1996).
\bibitem{WatChrGoe}T.~Watabe, Ch.~Christov, and
K.~Goeke, Bochum preprint RUB-TPII-3-96.
\bibitem{Witten} E. Witten, {\em Nucl.Phys.} {\bf  B223}, 433 (1983).
\bibitem{ANW} G.S. Adkins, C.R. Nappi, and E. Witten,
{\em Nucl. Phys.} {\bf B228}, 552 (1983).
\bibitem{DE}
D.Diakonov and M.Eides, {\it Sov. Phys. JETP Lett.} {\bf 38}, 433 (1983).
\bibitem{ATFR} I.J.R. Aitchison, C.M. Fraser,
{\em Phys. Lett.} {\bf 146B}, 63 (1984).
\bibitem{DSW}
A.Dhar, R.Shankar and S.Wadia, {\it Phys. Rev.} {\bf D31}, 3256 (1985).
\bibitem{DyPe1} D. Dyakonov and V. Petrov,
{\em Nucl.Phys.} {\bf B272}, 457 (1986); preprint LNPI-1153,
published in: {\em Hadron matter under extreme conditions}, p.192,
Kiew (1986).
\bibitem{DiPePo} D. Diakonov, V. Petrov, and P.V. Pobylitsa,
{\em Nucl. Phys}. {\bf B306}, 809 (1988).
\bibitem{Review} C.V. Christov, A. Blotz, H.-C. Kim,P. Pobylitsa,
T. Watabe, Th. Meissner, E. Ruiz Arriola and K. Goeke,
{\em Prog. Part. Nucl. Phys.} {\bf 37}, 91  (1996).
\bibitem{WakYos}
M.~Wakamatsu, and H.~Yoshiki, {\em Nucl. Phys}. {\bf A524}, 561 (1991).
\bibitem{Goeetal}
K.~Goeke, {\em et al.},
{\em Phys. Lett.} {\bf 256B}, 321 (1991).
\bibitem{ich} M. V. Polyakov,
{\em  Yad. Fiz.} {\bf 51}, 1110 (1990);\\
M. V. Polyakov, Diploma Thesis (in Russian), Leningrad (1989).
\bibitem{KaRi} S. Kahana and G. Ripka,
{\em Nucl.Phys. }{\bf A429}, 962 (1984).
\bibitem{PraBloGo1}
 M. Prasza\l owicz, A. Blotz and K. Goeke,
{\em  Phys. Rev.}
{\bf D47}, 1127 (1993).
\bibitem{Shiomi}
H. Shiomi,
{\em Nucl.Phys. }{\bf A603}, 281 (1996).
\bibitem{DoHo}
 J.F. Donoghue, B.R. Holstein,
{\em Phys. Rev.}{\bf D25}, 206 (1982).
\bibitem{Morita} M. Morita, R. Morita and K. Koshigiri,  {\em Nucl.Phys.}
{\bf A577}, 387c (1994).

\end{thebibliography}
\end{document}